\documentclass{article}
\usepackage[margin=1in]{geometry}
\usepackage{amsmath}
\usepackage{amssymb}
\usepackage{amsthm}
\usepackage{dcolumn}
\usepackage{mdwlist}
\usepackage{bm}
\usepackage{graphicx}
\usepackage{subfig}
\usepackage{cite}
\usepackage{calc}
\usepackage{xcolor}
\usepackage{hyperref}
\usepackage[capitalise]{cleveref}

\title{Imitation of Success Leads to Cost of Living Mediated Fairness\\ in the Ultimatum Game}

\author{Yunong Chen\footnote{
	Dept. of Mathematics,
    The Pennsylvania State University,
    University Park, PA 16802
    } \and Andrew Belmonte\footnotemark[1]~\footnote{
	Huck Institutes of the Life Sciences,
    The Pennsylvania State University,
    University Park, PA 16802
    }\and Christopher Griffin\footnotemark[1]~
\footnote{
	Applied Research Laboratory,
	The Pennsylvania State University,
    University Park, PA 16802
    }
}

\date{Preprint - \today}

\begin{document}
\maketitle

\begin{abstract} The mechanism behind the emergence of cooperation in both biological and social systems is currently not understood. In particular, human behavior in the Ultimatum game is almost always irrational, preferring mutualistic sharing strategies, while chimpanzees act rationally and selfishly. However, human behavior varies with geographic and cultural differences leading to distinct behaviors. In this paper, we analyze a social imitation model that incorporates internal energy caches (e.g., food/money savings), cost of living, death, and reproduction. We show that when imitation (and death) occurs, a natural correlation between selfishness and cost of living emerges. However, in all societies that do not collapse, non-Nash sharing strategies emerge as the de facto result of imitation. We explain these results by constructing a mean-field approximation of the internal energy cache informed by time-varying distributions extracted from experimental data. Results from a meta-analysis on geographically diverse ultimatum game studies in humans, show the proposed model captures some of the qualitative aspects of the real-world data and suggests further experimentation.
\end{abstract}

\section{Introduction}
Cooperation is critical for the emergence of societies (e.g., ants, cetaceans, humans etc.). However cooperation is frequently an irrational response to an environment with a cost of living. Consequently, understanding and modeling the mechanism of the emergence of cooperation and fairness is still an active area of research in social and biological theory \cite{A81,CZ97,JK01,NSTF04,SP05,HH06,SSP08,KI09,VPLS12,PG16,CWZC19,GDL20}. 
The \textit{Ultimatum Game} (UG) is an archetypal game illustrating both the difficulties in modeling concepts of fairness and cooperation. In the game, one player is given a sum of money which she must divide in some proportion between herself and a second player. The second player may then accept the offer, in which case the pot is divided accordingly, or reject the offer in which case each player receives nothing \cite{GBS95}. This is like a continuous variation of the Stag-Hunt game, in which individual gain competes against mutual benefit. The notional \textit{money} can act as a stand-in for a cooperative hunt, business venture etc. Here we introduce an additional UG variable, individual wealth, which drives the dynamic {\it imitate the successful}.

A considerable amount of theoretical and experimental research has been done on the ultimatum game (see e.g., \cite{BY98,SRAN03,OSK04,HF05,HMBE06,JCT07,YHTS09,B17,WCLJ07,CDFJ08,CAHS15}).
Classical game theory asserts the most rational, sub-game perfect solution is for the dividing player to keep as much of the prize as possible, while the deciding player accepts any offer. However, almost all experiments with humans (but not chimpanzees \cite{JCT07}) show that individuals will offer far more than the minimum quantity and deciding players will frequently reject offers at the expense of their own well-being (presumably as an act of punishment for unfair or non-cooperative behavior).  
In particular, Oosterbeek {\it et al.}~conducted a meta-analysis of 37 papers with 75 results from various countries \cite{OSK04}, and concluded that there is not a significant difference in proposers' behavior, but there is a difference in responders' behaviors across (geographic) regions. Any model of UG dynamics should include elements observed in their meta-study: 
(i) it must produce a diversity of results that can be tuned to explain geographic diversity; 
(ii) it should explain why offers show less variation than rejection rates; 
(iii) it should be generally consistent with human behavior.

Mathematical models by Nowak {\it et al.}~approach the Ultimatum Game from an evolutionary game theory perspective, by including the reputation of agents as part of the offer making process \cite{NPS00} and in a one-shot game context \cite{RTON13}. More recently, Gale {\it et al.}~construct a discrete strategy evolutionary game representation \cite{GBS95}, and show that evolutionary stable strategies exist in this game. In addition to this, a substantial amount of work has been done on spatial ultimatum games \cite{PNS00,PN01,KR08,IRS11}.

\section{Model}
Our proposed model is an agent-based simulation following the spirit of \cite{NPS00,RTON13}, similar to the approach taken in \cite{zhu2016,rajtmajer2017}. We introduce a dynamic wealth variable \cite{B14} for each player, as an integrated measure of success. In our model, agents interact randomly and each interaction is an instance of UG with a possible prize $P$. Agents are chosen at random to be the offerer or decider. The state of agent $i$ is specified by internal variables $(\lambda^{i},\theta^{i},B^{i})$ where $\lambda^{i} \in [0,1]$ is the fairness (cooperation) demanded by Agent $i$, $\theta^{i} \in [0,1]$ is the offer provided by Agent $i$ and $B^{i} > 0$ is the energy cache (bank) held by each agent, which keeps track of the winnings (energy) from each interaction. Energy loss in the system is set by the cost of living parameter $\kappa$, which is subtracted at each time step from the energy cache of each player.

If Agents $i$ and $j$ interact and $i$ is the offerer, then Agent $j$ rejects the offer whenever $\theta^{i} < \lambda^{j}$. In the case of acceptance, Agent $i$ keeps $(1 - \theta^{i}) P$ and Agent $j$ keeps $\theta^{i}P$. When $P = 1$, then all parameters can be expressed as ratios of $P$. Let $\chi_{ij}(t)$ be an indicator function that is $1$ at time $t$ exactly when $i$ and $j$ interact. The discrete time agent-based model dynamics are given by:
\begin{equation}
B^{i}(t + \epsilon) = B^{i} + \frac{\epsilon}{2}\sum_{j\neq i} \chi_{ij}(t)P\left(1-\theta^i\right)U(\theta^i-\lambda^j) 
+ \frac{\epsilon}{2}\sum_{j\neq i} \chi_{ij}(t) P\theta^jU\left(\theta^j-\lambda^i\right) - \kappa,
\label{eqn:BankSim}
\end{equation}
where
\begin{displaymath}
U(x) = \begin{cases}
1 & \text{if $x > 0$}\\
0 & \text{otherwise}
\end{cases}
\end{displaymath}
is the unit step function, and we take $\kappa \in [0,1]$.
For simulation purposes, we set $\epsilon = 1$. Taking the expected value of these equations, a mean-field approximation of the agent energy dynamics can be derived:
\begin{equation}
\Delta \hat{B}^i(t) = \frac{\epsilon}{2q}\sum_{j\neq i} \left[ \left(1-\theta^i\right)U\left(\theta^i-\lambda^j\right) \theta^jU\left(\theta^j-\lambda^i\right)\right] - \kappa,
\label{eqn:B}
\end{equation}
The normalizing value $q$ is given by:
\begin{displaymath}
q = \begin{cases}
n & \text{n is odd}\\
n-1 & \text{otherwise}
\end{cases},
\end{displaymath}
which models the random choice of two agents from a completely connected population of $n$ agents. We next propose dynamics that drive the population towards a statistical equilibrium $(\theta^i,\lambda^i) \to (\lambda^*,\theta^*)$. However, independent of any game dynamics for the population, we can already derive certain relations that characterize the dynamics of the energy cache $B$ using \cref{eqn:B}. If $\lambda^* > \theta^*$, then $U(\theta^* - \lambda^*) = 0$ and $\Delta \hat{B} < 0$. Populations of this type will collapse. On the other hand, if $\lambda^* < \theta^*$, then as $n \to\infty$:
\begin{equation}
\Delta{\hat{B}}(t) =\frac{\epsilon}{2} \left(1 - 2\kappa\right),
\label{eqn:DeltaB}
\end{equation}
which also holds in general for even $n$. Thus, if $\kappa > \tfrac{1}{2}$, the population will collapse in the mean. For $\kappa = \tfrac{1}{2}$, the population energy caches will stabilize in the mean and for $\kappa < \tfrac{1}{2}$, the population energy caches will increase without bound. 


In discrete time, the dynamics of $(\theta,\lambda)$ are given by:
\begin{align}
\lambda^i(t+\epsilon) &= \lambda^i(t) + \epsilon\sum_{j \neq i} (\lambda^j - \lambda^i) p_{ij}\label{eqn:Lambda}\\
\theta^i(t+\epsilon) &= \theta^i(t) + \epsilon\sum_{j \neq i} (\theta^j - \theta^i)p_{ij},\label{eqn:Theta}
\end{align}
where $p_{ij}$ are imitation probabilities. Let:
\begin{equation}
Q^i = \sum_{j}U(B^j - B^i),
\end{equation}
this is the cumulative difference in energy values for all agents $j$ with $B^j > B^i$. For the discrete time simulation, we set:
\begin{equation}
p_{ij} = \begin{cases}
\frac{ (B^j - B^i) U(B^j - B^i) }{\sum_{h}(B^h - B^i)U(B^h - B^i)} & \text{if $Q^i > 0$}\\
0 & \text{otherwise}
\end{cases}
\label{eqn:DiscreteImitation}
\end{equation}
\cref{eqn:Lambda,eqn:Theta} are imitation dynamics in which agents imitate those who outperform them. Thus Agent $j$ does not rationally choose $(\lambda^j,\theta^j)$, but adjusts these values based on observations weighted towards other, more successful agents. Our model is based on recent research showing that children will imitate higher status individuals more selectively than lower status individuals \cite{M13}. Additionally, children will infer status based on observing imitation in adults \cite{OC15}. Lastly, \cite{GL20} shows that in strategic settings humans will imitate behavior based on pay-off inequality.

For imitation systems like \cref{eqn:Lambda,eqn:Theta}, Griffin \textit{et al.} proved that a sufficient condition for convergence is the emergence of a fixed leader $i^*$ imitated (directly or indirectly) by all agents \cite{GRSB19}, which readily occurs in this system as a result of the total ordering of $B^i$. As $\epsilon \to 0$, \cref{eqn:Lambda,eqn:Theta} become the continuous time consensus equations as surveyed in \cite{MT14}, but with a state-varying coefficients. The proof of convergence in \cite{GRSB19} for discrete time updates suggests that exact values of $p_{ij}$ are irrelevant, as long as Agent $i$ is imitating those agents who outperform it. 

Whether in continuous or discrete dynamics, these systems have an infinite set of fixed points $\theta^k = \theta^*$, $\lambda^k = \lambda^*$ for $\theta^*,\lambda^* \in [0,1]\times[0,1]$. It is clear that not all such fixed points are equally likely or even realistic, since all systems with $\lambda^* > \theta^*$ would lead to population collapse for any cost of living $\kappa > 0$. Therefore, the distributions of long-run behavior in these systems should provide  insights into the emergence of cooperative or fair behaviors.

We assume agents are initialized with $\theta^k$ and $\lambda^k$ uniformly distributed in $[0,1]$. Again consider the case when $n \to \infty$. From \cref{eqn:B} the expected per-round energy increase near $t = 0$ for an agent with parameters $(\theta,\lambda)$ is:
\begin{equation}
\Delta B(\theta,\lambda,\kappa) = \frac{1}{2}\int_0^\theta (1-\theta)\, d\lambda + \frac{1}{2}\int_\lambda^1 \theta\,d\theta - \kappa = 
\frac{1}{2} (1-\theta ) \theta +\frac{1}{2} \left(\frac{1}{2}-\frac{\lambda
   ^2}{2}\right) -\kappa.
\label{eqn:ProbEval}
\end{equation}
Maximizing this expression subject to the constraints $0\leq\theta,\lambda\leq 1$, suggests the optimal fairness demand is $\lambda^+ = 0$, while the best offer is $\theta^+ = \tfrac{1}{2}$. This is consistent with the classical Nash equilibrium ($\lambda^+ = 0$) but also consistent with fairness considerations ($\theta^+ = \tfrac{1}{2}$), since an agent can never be certain whether she will interact with an agent with high or low $\lambda$. If the players were perfectly rational, then a true Nash equilibrium would be $\theta^{\text{NE}} = \lambda^{\text{NE}} = 0$, since rational players realizing $\lambda^+ = \lambda^{\text{NE}} = 0$ would make $\theta^{\text{NE}} = 0$. Our empirical results show that this equilibrium does not result from imitation. 

From \cref{eqn:ProbEval}, when $\kappa > \tfrac{3}{8}$, the expected increase for even an optimal player is negative. This will lead to a mean decrease in energy caches until imitation leads to higher success rates in UG.
Let $\chi_{\Delta B}(\theta,\lambda,\kappa)$ be an indicator function that is 1 just in case, \cref{eqn:ProbEval} is positive. Numerical evaluation shows that when $\kappa^* \approx 0.26246$:
\begin{displaymath}
\int_0^1\int_0^1\chi_{\Delta B}(\theta,\lambda,\kappa^*)\,d\lambda\,d\theta = \tfrac{1}{2}.
\end{displaymath}
For $\kappa > \kappa^*$, the median energy cache value will decrease in early interactions before imitation can contract the strategy space.
Individuals whose energy cache reaches zero are assumed dead and can no longer interact in the system. Reproduction or replacement of players is used to maintain a constant population, and the specific rule we use is described in the simulation details below. 

\section{Simulation Results}
We simulate a population with $N$ agents. Agents are initialized with an energy cache value $B^i$, and uniformly randomly assigned values $\theta^i$ and $\lambda^i$. Agents enter a \textit{game loop}, where each agent plays UG with another randomly selected agent. Once all agents have played, energy caches are updated accordingly. In the agent-based simulation, we introduce a reproductive step into the mimicking process to account for agents with non-positive energy cache and to identify population collapse prior to convergence. If all agents have $B^i < 0$ after subtracting the cost of living, the simulation ends immediately. Otherwise, all agents with $B^i > 0$, mimic others using \cref{eqn:Lambda,eqn:Theta} in a \textit{mimic/reproduce} loop. If all agents have survived, agents return to the game loop. Otherwise, agents are randomly chosen to reproduce with probability proportional to their energy cache; i.e., the fittest reproduce with higher probability. Reproduction continues until the population reaches $N$. If the population never collapses, the process is terminated after $T$ rounds. The size of $T$ is chosen to ensure convergence. To ensure numerical validity, the model was implemented both in Python and Mathematica, and results were compared to ensure statistical consistency. 

\begin{figure}[htbp] 
\centering
\includegraphics[width=0.35\columnwidth]{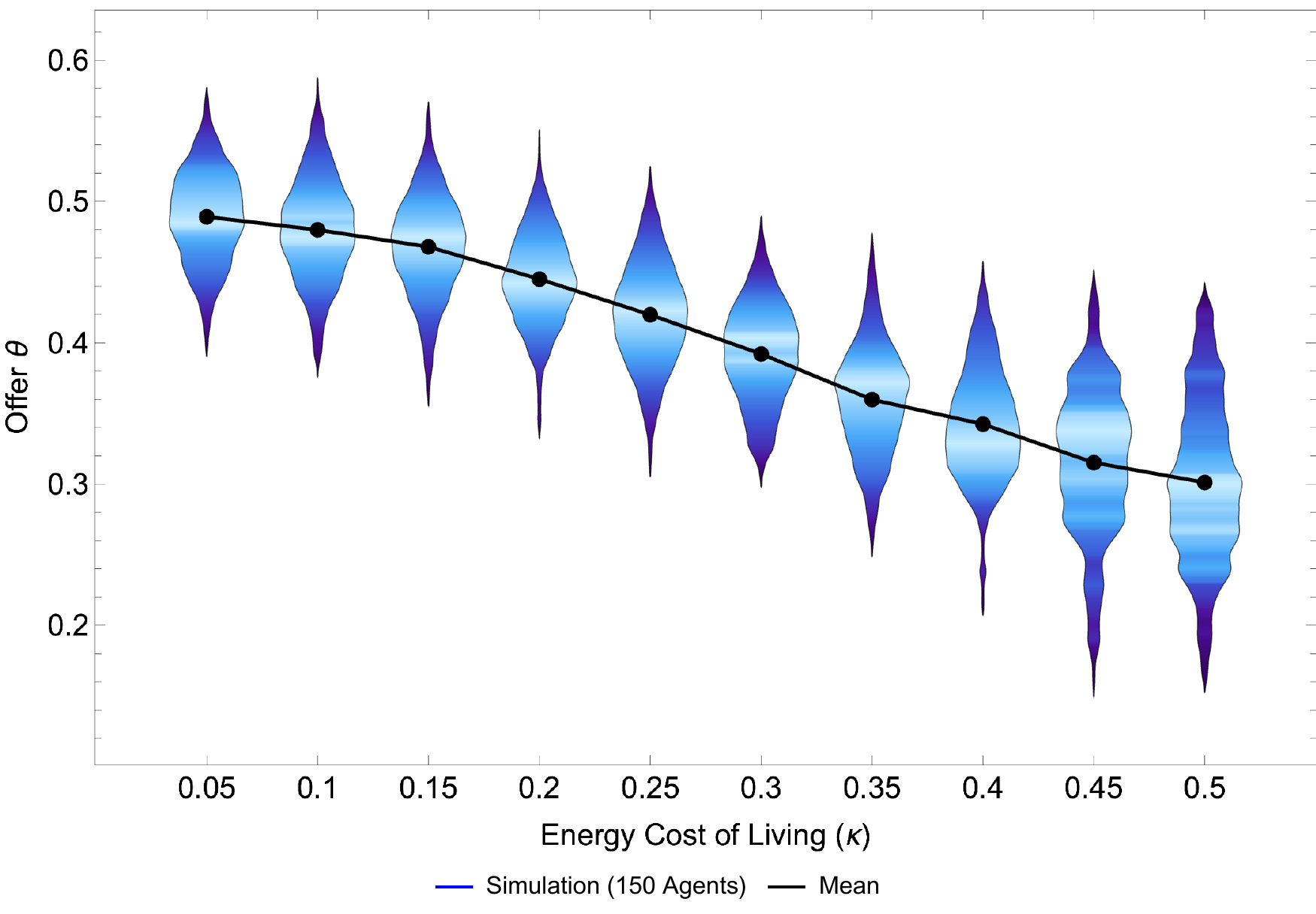}\\
\label{fig:Fairness}\includegraphics[width=0.35\columnwidth]{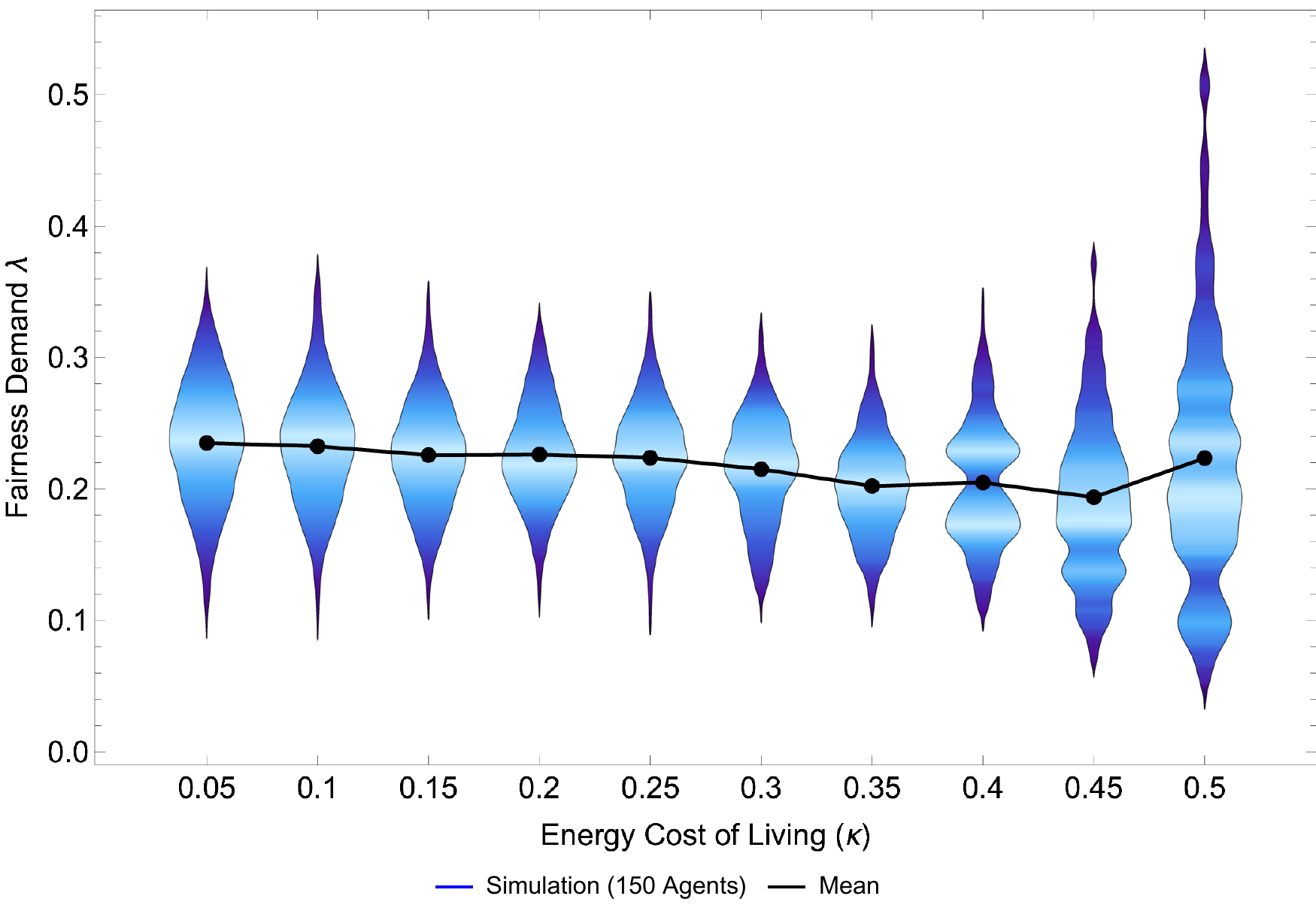}\\
\includegraphics[width=0.35\columnwidth]{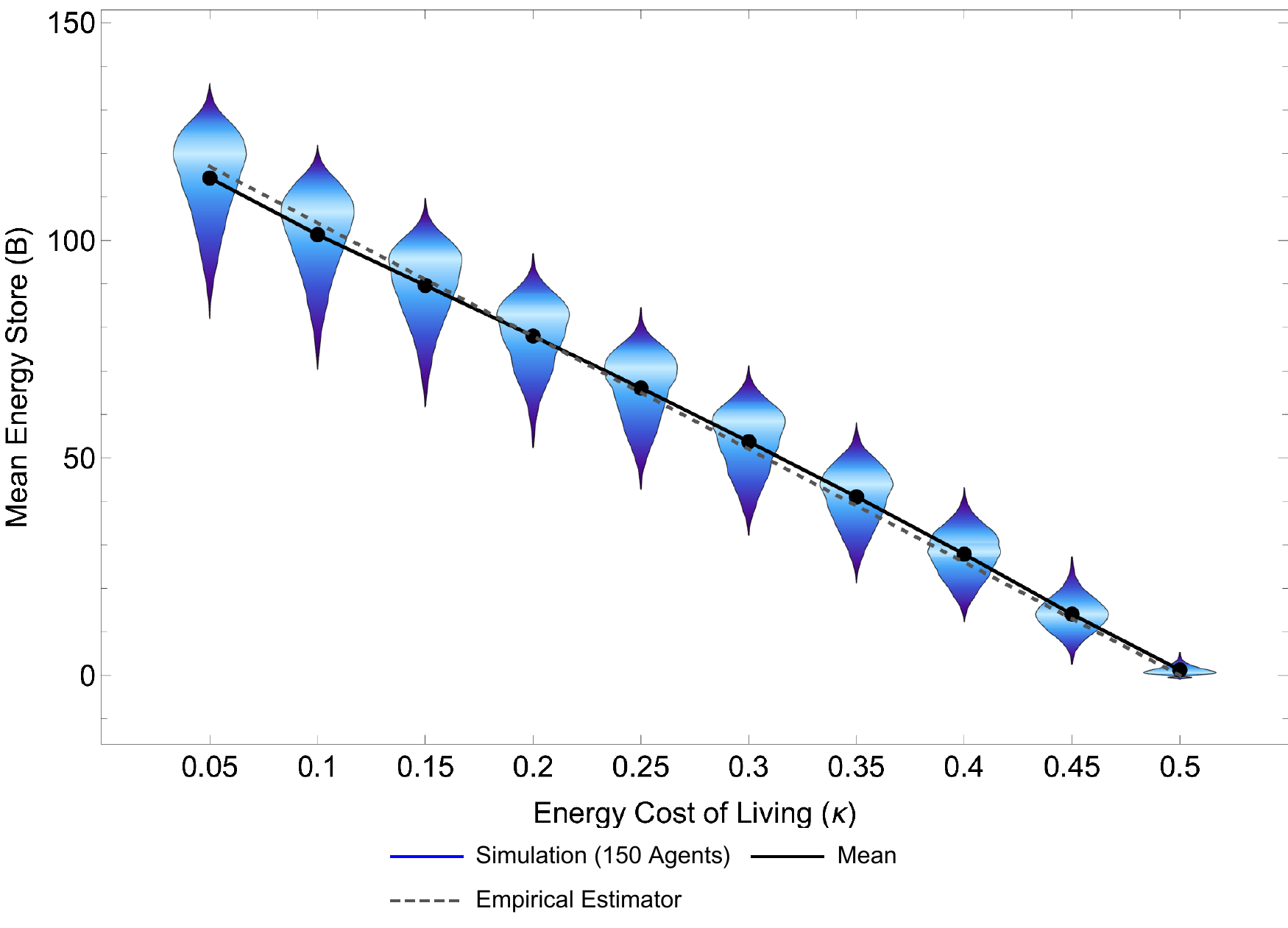}
\caption{Simulation results using 150 agents, 300 rounds of play, and 100 replications: (top) there is a well defined negative correlation between cost of living and offer; 
(middle) Fairness demands are stable across cost of living values; 
(bottom) energy cache values follow an empirical trend derivable from the model.
}
\label{fig:Results1}
\end{figure}

\begin{figure*}[t]
\centering
\includegraphics[width=0.98\textwidth]{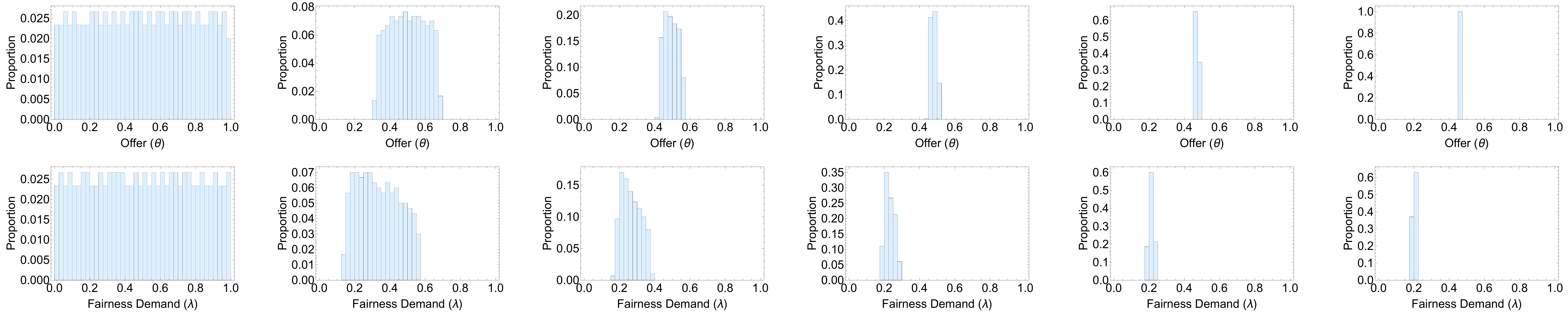}
\caption{(top) Convergence of the distribution of $\theta^i$ from a uniform distribution to a delta distribution. (bottom) Convergence of the distribution of $\lambda^i$ from a uniform distribution to a delta distribution. (both) $\kappa = 0.1$, 300 agents are simulated. Times go from $t = 0$ to $t = 500$ in increments of $\Delta t = 100$. }
\label{fig:Convergence}
\end{figure*}

\begin{figure}[htbp] 
\centering
\includegraphics[width=0.35\columnwidth]{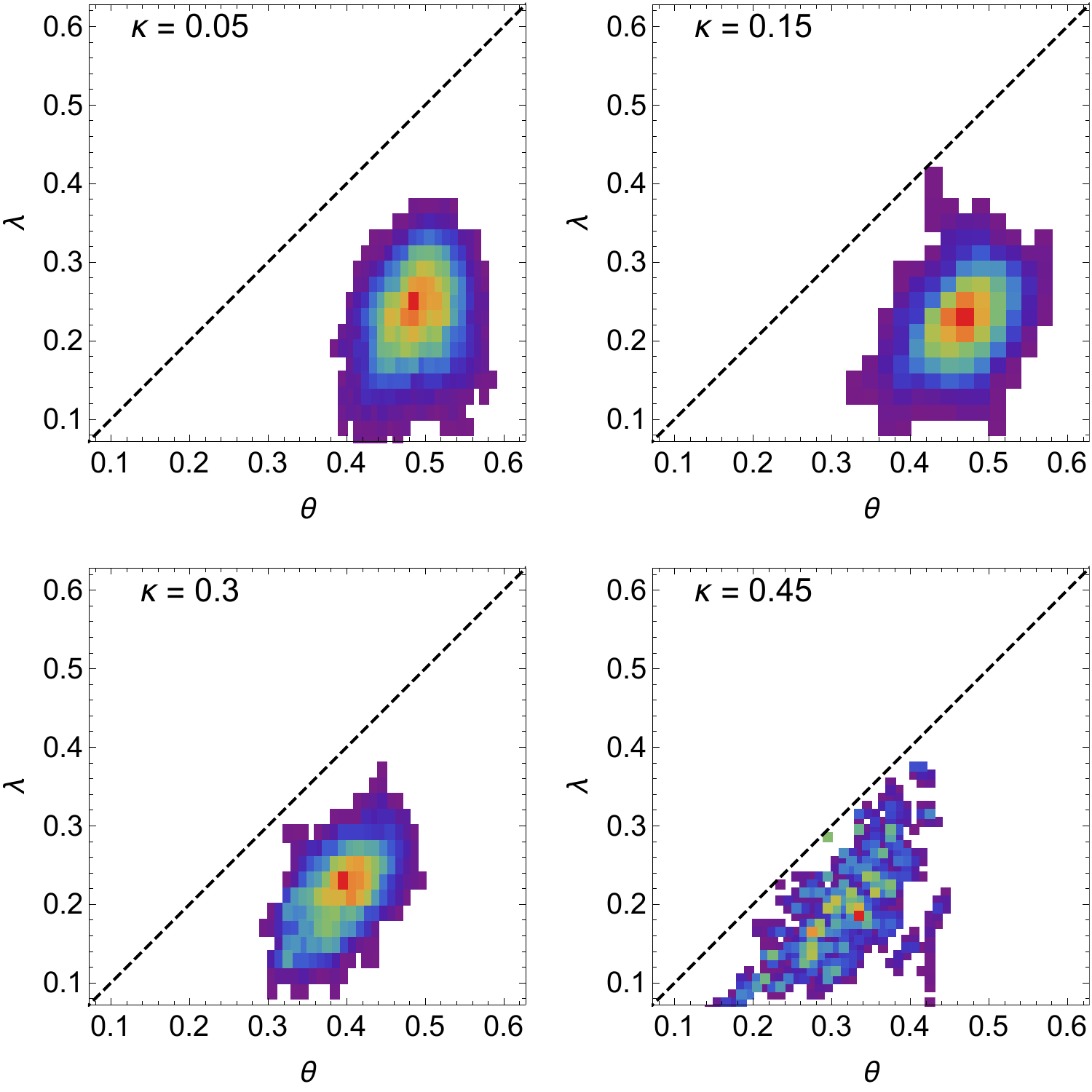}
\caption{Density plots show the distributions of $(\theta^*,\lambda^*)$ over multiple replications with varying costs of living.}
\label{fig:DensityPlot}
\end{figure}

\cref{fig:Results1} shows simulation results for $N = 150$ players and running time $T = 300$. All agent energy caches are initially set to $1$. We used 100 realizations (replications). Distribution plots for $B$, $\theta$ and $\lambda$ are shown, with cost of living $\kappa$ ranging from $0.05$ to $0.5$. 
Density plots showing the joint converged ($\theta$, $\lambda$) distributions are shown in \cref{fig:DensityPlot}.
The convergence of $\theta^i(t)$ and $\lambda^i(t)$ is illustrated in \cref{fig:Convergence} for $300$ agents, $T = 500$ and $\kappa = 0.1$. To create this figure, 100 replications were constructed and $\theta^i(t)$ and $\lambda^i(t)$ were sorted at each round. These sorted lists where then averaged (over replication) to obtain $\bar{\theta}^{[i]}(t)$ and $\bar{\lambda}^{[i]}(t)$, where $\bar{\theta}^{[i]}(t)$ is the mean offer value of the agent with the $i^\text{th}$ smallest offer value. The quantity $\bar{\lambda}^{[i]}(t)$ is defined analogously.

The simulation shows downward pressure on the offer value correlated with the energy cost of living $\kappa$ with consistent values of $\lambda^*$ between (approximately) $0.1$ and $0.4$. As is expected, the value of $\hat{B}^i(t)$, the mean energy store value decreases as a function of cost of living. We derive an empirical linear approximation for the mean, which we discuss in the sequel. Understanding the origin of this relationship is complicated by the fact that there is no convenience closed form expression for $\lambda^i(t)$ or $\theta^i(t)$. To remedy this, we use a combination of empirical distribution modeling and closed form analysis of $\Delta\hat{B}^i$ to explain the observed behavior. 

\subsection{Mixed Empirical and Closed form System Modeling}
\begin{figure}[htbp] 
\centering
\includegraphics[width=0.35\columnwidth]{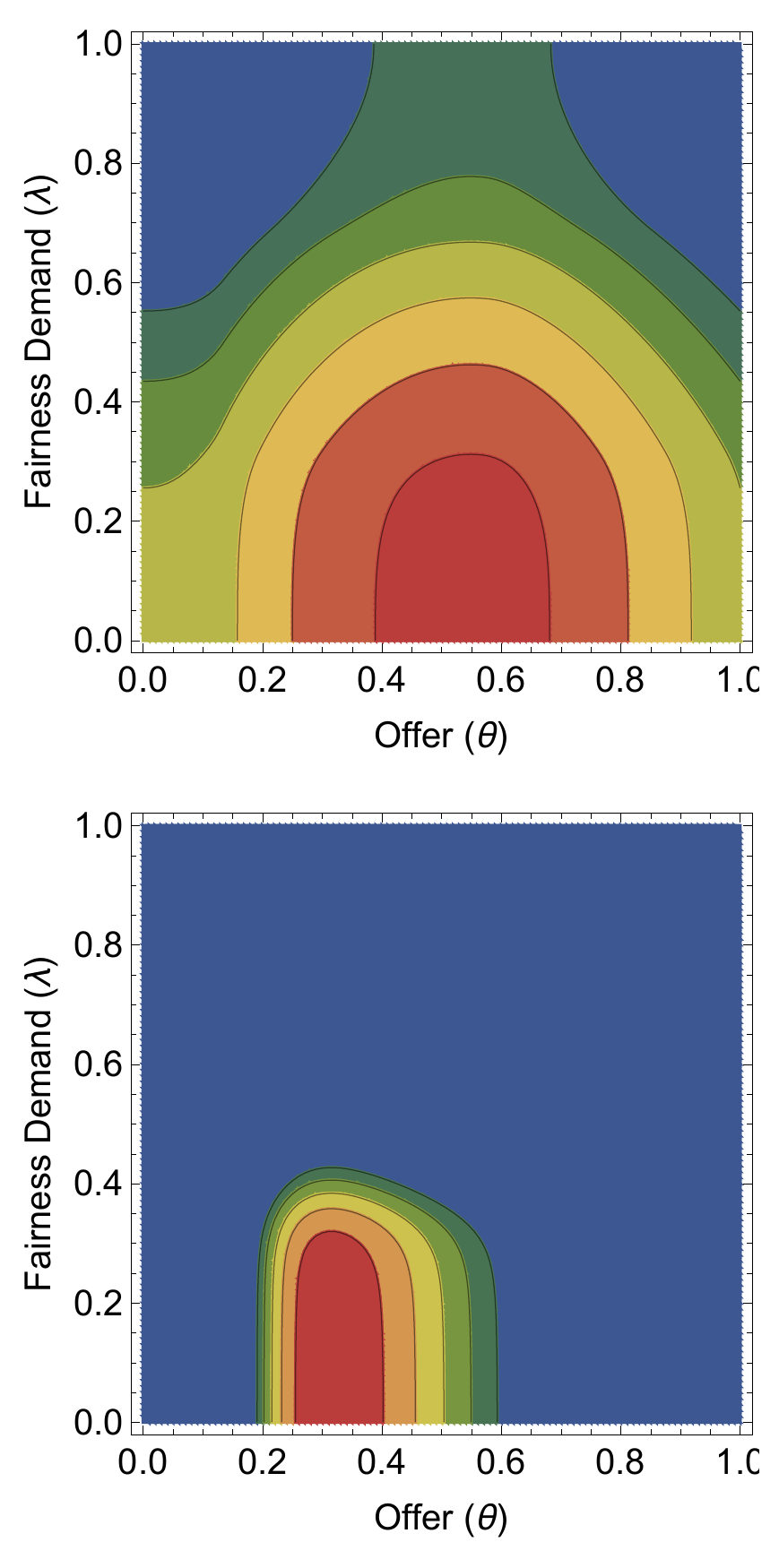}
\caption{(top) (bottom)}
\label{fig:Asymmetry}
\end{figure}

At arbitrary time $t$ when the distribution of fairness demands and offers is given by probability density functions $f^t_\lambda(s)$ and $f^t_\theta(s)$ respectively, then \cref{eqn:ProbEval} can be generalized as:
\begin{equation}
\Delta \hat{B}(t;\theta,\lambda,\kappa) = 
\frac{1}{2}\left(\int_{0}^\theta (1-\theta)f_\lambda^t(s) \,ds + 
\int_\lambda^1 sf_\theta^t(s)\,ds 
\right)- \kappa
\label{eqn:DBGeneral}
\end{equation}
This expression cannot be computed without the time-varying distributions in question, which cannot be computed without an appropriate Fokker-Plank equation, which is difficult to construct. To compensate, we can fit distributions to the data $\bar{\lambda}^{[i]}(t)$ and $\bar{\theta}^{[i]}(t)$ to obtain estimators $\hat{f}^t_\lambda(s)$ and $\hat{f}^t_\theta(s)$, which can be used in \cref{eqn:DBGeneral}. These empirically estimated distributions stand in for the mean-field distributions. Not: All distributions were estimated using Mathematica's \texttt{FindDistribution} function. We can then compute:
\begin{equation}
\hat{B}(t;\theta,\lambda,\kappa) = 
\begin{cases}
B_0 & \text{if t = 0}\\
\sum_{s\leq t}\Delta{B}(t;\theta,\lambda,\kappa)U[B(t-1;\theta,\lambda,\kappa)] & \text{otherwise}
\end{cases},
\label{eqn:BEst}
\end{equation}
where the factor $U[B(t-1;\theta,\lambda,\kappa)]$ sets $\hat{B}(t;\theta,\lambda,\kappa) = 0$ if $\hat{B}(t-1;\theta,\lambda,\kappa) = 0$. That is, it models the death of a test agent with parameters $(\theta,\lambda)$. The imitation dynamics defined by \cref{eqn:Lambda,eqn:Theta,eqn:DiscreteImitation} imply that the larger $\hat{B}(t;\theta,\lambda,\kappa)$ the more likely an agent with parameters $(\theta,\lambda)$ will be imitated. Thus, we can use $\hat{B}(t;\theta,\lambda,\kappa)$ at an appropriately large time (e.g., $t = 100$) to estimate which agents are most likely to be imitated for a given $\kappa$. We show this estimation for $\kappa = 0.1$ and $\kappa = 0.4$ in \cref{fig:Asymmetry}.
To test this model, we use the top 5\% of computed values of $\hat{B}(t,\theta,\lambda,\kappa)$ to compute estimated intervals on the values of $\theta^*$ and $\lambda^*$ for $\kappa = 0.1$ and $\kappa = 0.4$. We compare these intervals with the $5\% - 95\%$ intervals computed from the experimental results shown in \cref{fig:Results1}. This is shown in \cref{tab:Estimate}.
\begin{table}[htbp]
\centering
\begin{tabular}{|l|c|c|}
\hline
$\kappa$ & Model Est. Interval & Computed Interval\\
\hline
$0.1$ & $[0.43,0.64]$ & [0.42,0.54]\\
\hline
$0.4$ & $[0.25, 0.41]$ & [0.280.41]\\
\hline
\end{tabular}\\
\textbf{Offer Estimate}

\vspace*{1em}

\begin{tabular}{|l|c|c|}
\hline
$\kappa$ & Model Est. Interval & Computed Interval\\
\hline
$0.1$ & $[0,0.33]$ & [0.15,0.31]\\
\hline
$0.4$ & $[0, 0.26]$ & [0.13,0.28]\\
\hline
\end{tabular}\\
\textbf{Fairness Demand Estimate}
\caption{Comparison of estimated and computed intervals on $\theta^*$ and $\lambda^*$ using information from \cref{eqn:BEst}.}
\label{tab:Estimate}
\end{table}
These results are both consistent with and predictive of the distributions seen in \cref{fig:Results1}; i.e., they explains both the downward slope of $\theta$ as a function of $\kappa$ in  \cref{fig:Results1} (top) and the relatively constant behavior of $\lambda$ as a function of $\kappa$. We stress that estimations in \cref{fig:Asymmetry} and \cref{tab:Estimate} are generated by a model (\cref{eqn:DBGeneral} and \cref{eqn:BEst}) with distribution constants determined empirically. Thus an area of future work is to replace these empirically determined distributions with modeled distributions. 

\subsection{Asymptotic Behavior of $\hat{B}$}
The dynamics of the energy cache values can be modeled asymptotically. As $t \to \infty$, $f^t_\lambda(s) \approx \delta(s-\lambda^*)$ and $f^t_\theta(s) \approx \delta(s-\theta^*)$, where $(\theta^*,\lambda^*)$ is the fixed point of the $(\theta(t),\lambda(t))$. This is illustrated in \cref{fig:Convergence}. As $t \to \infty$, the energy caches of each agent asymptotically approaches:
\begin{displaymath}
 B^i(t) = (\tfrac{1}{2}-\kappa)t.
\end{displaymath}
This model is shown in \cref{fig:Results1} (bottom).
This over-estimates the long-run energy cache value because of the initial time taken to converge. We can approximate the trend seen in \cref{fig:Results1} (bottom), by noting that the time for $\lambda^i(t)$ and $\theta^i(t)$ to converge so that most UG interactions are successful in approximately 80 rounds (out of the 300 rounds simulated). Assuming that prior to convergence, only half of all interactions result in a successful UG, we obtain a thermodynamic-type relationship between the mean wealth of the population and the cost of living:
\begin{equation}
\tilde{B}(\kappa) = 260 \left(\frac{1}{2}-\kappa \right),
\label{eqn:Btilde}
\end{equation}
which explains the linear decrease with $\kappa$ shown in \cref{fig:Results1} (bottom), where we show the fit of \cref{eqn:Btilde}.

\begin{figure}[htbp]
\centering
\includegraphics[width=0.35\columnwidth]{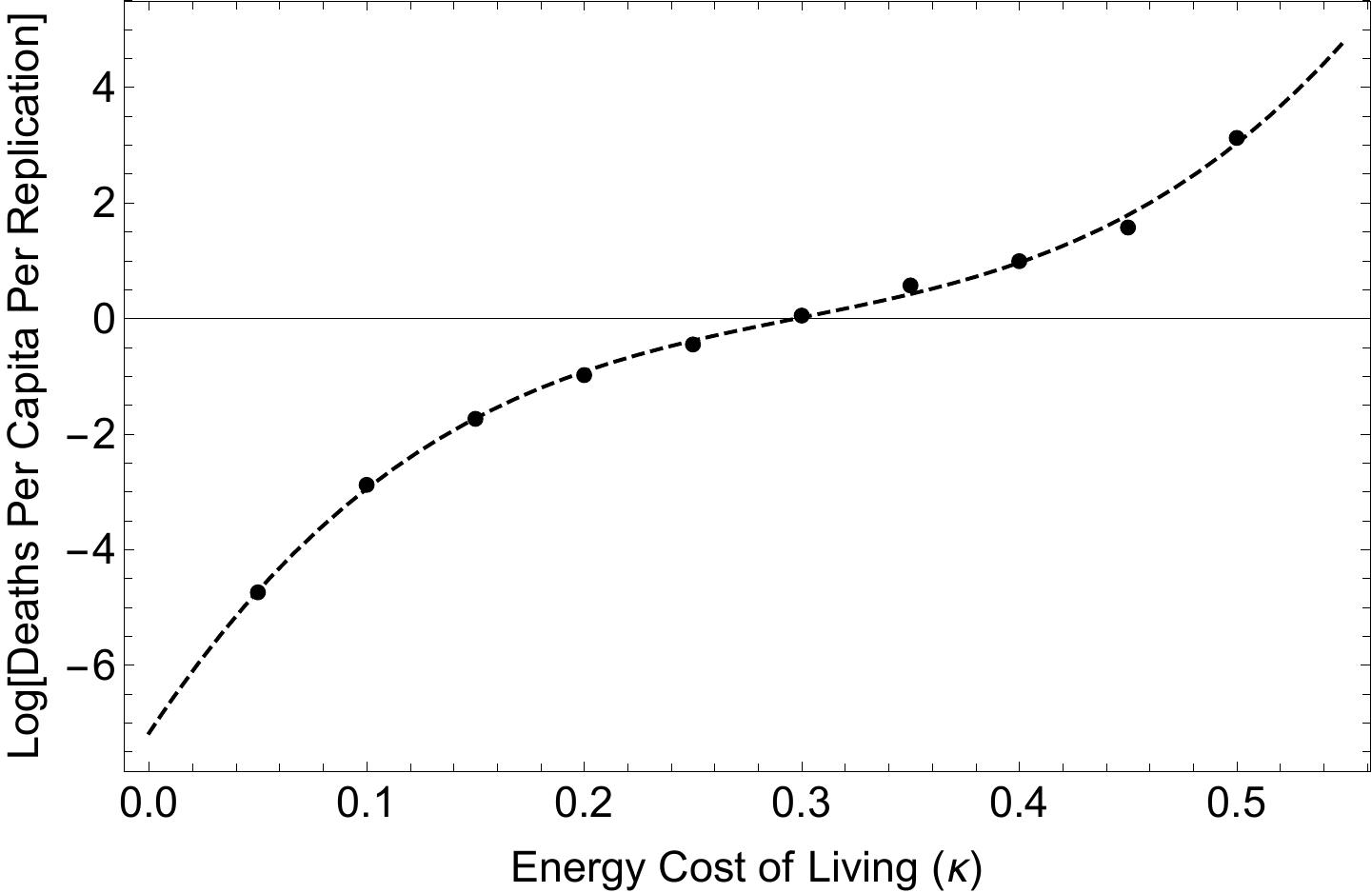}\\
\caption{Log-Log plot of per capita death rate in the simulation as a function of energy cost of living. 
}
\label{fig:Death}
\end{figure}

Figure \ref{fig:Death} shows the a log-plot of mean deaths per capita per simulation with notional cubic fit. As expected, there is a non-linear jump for $\kappa = \tfrac{1}{2}$. In addition, we note tht the zero-crossing (corresponding to one death per round) occurs roughly at $\kappa = \kappa^*$, representing a transition in the population to a more rapidly increasing per capita death rate with cost of living $\kappa$. This also correlates with more than 50\% of the agents having initially decreasing energy caches, thus increasing the per capita death rate.  

The global dynamics displayed in \cref{fig:Results1} are robust to changes in the speed of the underlying dynamics. In particular, we tested models in which (i) we replaced the discrete dynamics with continuous time differential equations (by letting $\epsilon \to 0$), (ii) an Euler step approximation of the resulting differential equations and (iii) a heterogenous starting energy cache value with the previously described dynamics. The ODE variants model fast imitation (on the time scale of the game play). In all cases except one, we included reproduction as a hybrid step by solving the ODEs for short time horizons, checking for death and then restarting the ODEs from the previous condition after removing agents with $B^k < 0$. All models used 100 replications except when reproduction was eliminated in which case 200 replications were used to ensure  statistically signifiant sample sizes. (Samples with population collapse were discarded.)

\begin{figure}[!ht]
\centering
\subfloat[Offer]{\includegraphics[width=0.35\columnwidth]{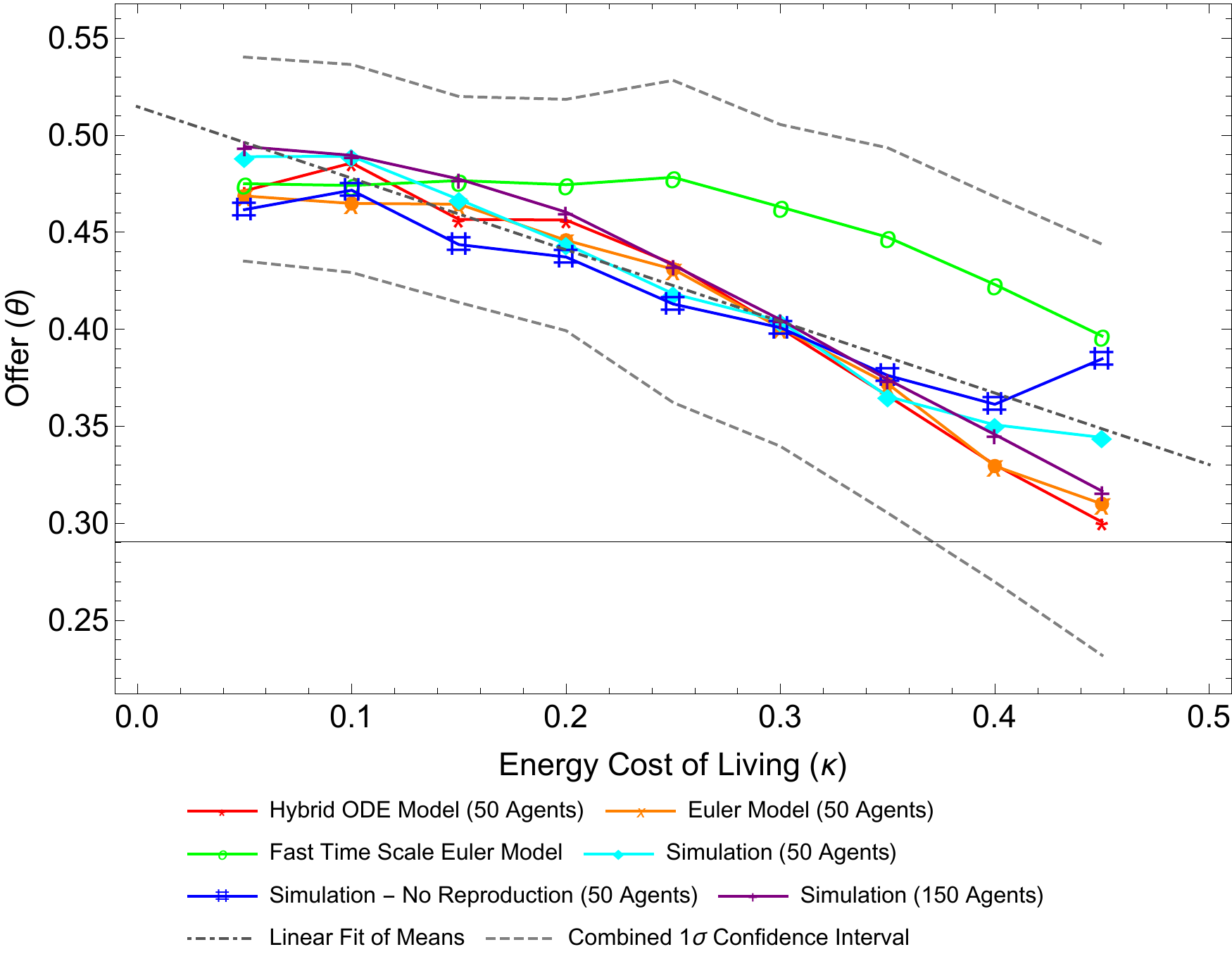}}\\
\subfloat[Fairness Demand]{\includegraphics[width=0.35\columnwidth]{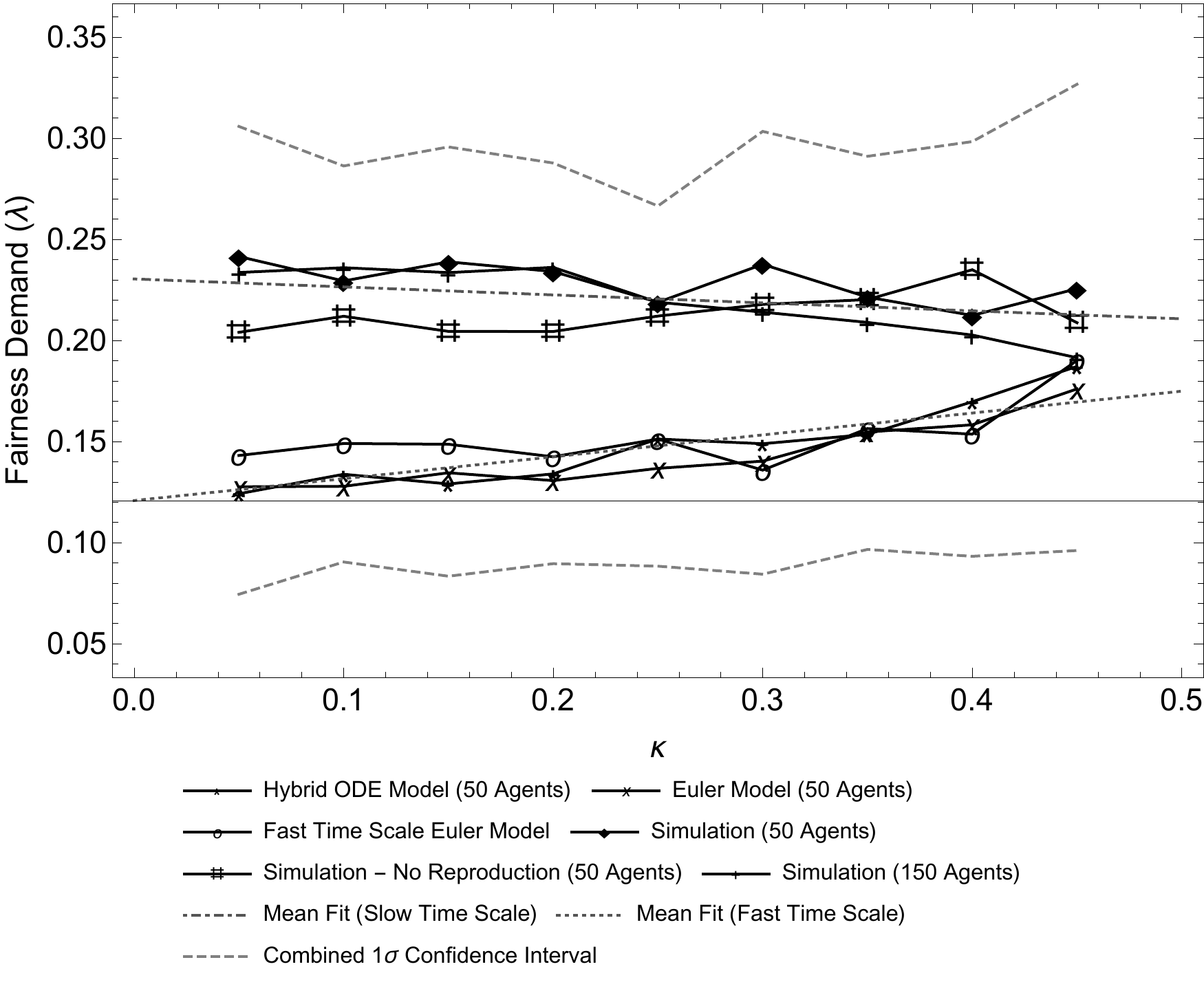}}\\
\caption{(top) Model variations show the same trend in offer as a function of cost of living.  (middle) Model variations show differing trend in fairness demand depending on imitation speed. (bottom) Scaling GDP as a proxy for cost of living in real-world data shows good correlation with the proposed model (number in parentheses denotes the number of countries averaged in each data point \protect \cite{OSK04}).}
\label{fig:Experiments}
\end{figure}

Results from robustness experiments are shown in \cref{fig:Experiments} (top), where we show the mean values $\bar{\theta}^*$. The envelopes are $1\sigma$. Similar tests were run for $\bar{\lambda}^*$ -- see \cref{fig:Experiments} (middle).
For all cases, $\bar{\theta}$ is decreasing in $\kappa$. The mean fit line has negative slope as a function of $\kappa$ ($p = 5.1\times 10^{-6}$) and adjusted $r^2$ of $0.95$, consistent with prior results and theoretical analysis. There is a difference in the behavior of $\bar{\lambda}^*$ for the the discrete time simulations and the continuous time (hybrid) variations. In the case of the hybrid ODE models (with or without Euler step approximations) $\bar{\lambda}^*$ increases as a function of $\kappa$ ($p < 0.002$) while for discrete step simulations $\bar{\lambda}^*$ decreases as a function of $\kappa$ ($p < 0.002$). When the data are combined, $\bar{\lambda}^*$ increases as a function of $\kappa$ but with $p < 0.004$, suggesting this effect may disappear with larger samples. This would be the expected behavior as indicated by \cref{fig:Asymmetry}.

\section{Discussion}
The proposed model provides behavior consistent with observations in the meta-study by Oosterbeek {\it et al.} \cite{OSK04} insofar as a diversity of offer proportions and rejection rates are shown to be possible as a result of random interaction and imitation of multiple agents. Over all simulations, the grand mean $\langle{\theta}\rangle = 0.43\pm0.00032$, which is consistent with the $40.41\%$ offer rate observed in \cite{OSK04}. The grand mean $\langle{\lambda}\rangle = 0.20\pm 0.00026$, implying that $20\%$ of offers would be rejected if chosen uniformly randomly from $[0,1]$. \cite{OSK04} reports a mean rejection rate of $16\%$, which is consistent with the results produced by the model assuming some cultural learning (imitation) has occurred. 
As noted, the distributions describing $\lambda^*$ have ranges from approximately $0.1$ to $0.4$, adequately modeling the large variation in rejection rates. Despite these similarities, we cannot fully validate the model empirically because neither \cite{OSK04} nor its constituent studies include a variable like cost of living. Their study does regress against GDP and reward as a percent of per capita GDP (which spans 2 orders of magnitude), but this is not an accurate measurement of intrinsic cost of living, especially in geographically diverse areas like the United States. We note that in \cite{OSK04}, regression of offer against GDP shows an insignificant negative correlation, which is consistent with \cref{fig:Results1} and \cref{fig:Experiments} (top), but these studies were not designed to measure this relationship. The results of this study suggest potential experimental analysis that could be done in controlled laboratory settings. 

\section{Conclusion}
Game Theory finds application in biological and social sciences, yet well-known occurrences like cooperation and altruism remain challenging within its rational self-interest assumptions. Our paper presents a novel approach to the canonical Ultimatum Game (UG), introducing an additional savings variable (energy) along with a cost of living. In our nonlinear agent-based model, energy represents success and drives imitation. Agents evolve toward fair sharing, but are more selfish with higher costs of living, with consistently lower fairness demands of others. This behavior is explained and predicted using a model with empirical determined distribution parameters. The model reproduces some empirical data of human UG performance across cultures, providing a new theoretical framework for heterogenous cooperation among humans. In future work, we will explore these dynamics further to determine whether the exact structure of the distributions on $\theta$ and $\lambda$ can be determined. This would remove the need to fit the distributions as a part of the modeling process and provide a complete mean-field dynamics for this system. 

\section*{Acknowledgement} CG and AB were supported in part by the National Science Foundation under grant DMS-1814876. The authors would like to thank S. Rajtmajer for her feedback on earlier drafts. CG thanks R. Bailey (USN) who ran the very first simulation of this phenomena in MATLAB while at the United States Naval Academy.

\bibliographystyle{unsrt}

\bibliography{../UltimatumGame-PRL/UltimatumGame}
\end{document}